\documentclass[aip,apl,reprint]{revtex4-2} 
\usepackage{graphicx} 
\usepackage{amsmath} 
\usepackage{textcomp, gensymb}
\usepackage[hidelinks,colorlinks=true]{hyperref} 
\usepackage{siunitx} 
\usepackage{orcidlink} 
\usepackage{subcaption}
\begin{document}

\title{Reduced decay in Josephson coupling across ferromagnetic junctions with spin-orbit coupling layers}

\author{Ivan Kindiak\,\orcidlink{0009-0006-8501-5341}}
\author{Swapna Sindhu Mishra\,\orcidlink{0000-0003-4074-4795}}
\author{Andrea Migliorini\,\orcidlink{0000-0002-6904-3573}}
\author{Banabir Pal\,\orcidlink{0000-0003-2423-4373}}
\author{Stuart S.P. Parkin\,\orcidlink{0000-0003-4702-6139}}
\email{stuart.parkin@mpi-halle.mpg.de}
\affiliation{Max Planck Institute of Microstructure Physics, Weinberg 2, 06120 Halle, Germany}

\date{\today}

\begin{abstract}

The generation of $S_z=1$ triplet Cooper pairs has been predicted theoretically in superconducting-ferromagnetic hybrid heterostructures in the presence of spin-orbit coupling.\cite{Bergeret_2014, Jacobsen2016Apr} In this study, we experimentally investigate vertical Josephson junctions where the weak link is formed from a ferromagnetic layer with perpendicular magnetic anisotropy sandwiched by two non-magnetic layers with weak or strong spin-orbit coupling. We find that the decay of the Josephson coupling is reduced in the latter case, possibly indicating the presence of $S_z=1$ spin-triplet correlations. We speculate that the canted magnetization required for these correlations is provided by the interaction of magnetization with Meissner effect in the superconducting layers.

\end{abstract}

\maketitle 

The superconducting proximity effect at interfaces between conventional superconductor (SC) and ferromagnetic (FM) layers enables unconventional Cooper pairing such as odd-frequency correlations which may carry a non-zero total spin. \cite{Bergeret_2001, Buzdin_2005,Linder2019Dec} Such systems have sparked a significant interest within the research community in recent years, as they may give rise to dissipationless spin-polarized currents. The effect of magnetization on supercurrents has been studied extensively and has found applications in cryogenic memory \cite{Ryazanov_2012, Soloviev_2017, Dayton_2018} and quantum technologies.\cite{Ioffe_1999, Blatter_2001, Yamashita_2005, Feofanov_2010} The spin-polarization induced in Cooper pairs  by passage through magnetic layers is governed by a combination of spin-mixing and spin-rotation effects.\cite{Eschrig2011Jan} This process is illustrated by the example of a non-collinear (NC) magnetic structure that is proximitized to a SC state. When a spin-singlet Cooper pair diffuses from a SC into a FM it experiences an exchange interaction and acquires a finite momentum.\cite{Demler_1997, Buzdin_2005} This leads to the spatial oscillation of the singlet ground state wave function, known as $0-\pi$ phase oscillations. Along with such short-range singlet oscillations, this process generates short-range triplet correlations (SRTCs) with $S_z=0$ which also oscillate spatially. In a quantization axis orthogonal to the initial magnetization direction at the SC/FM interface, the SRTCs are `rotated' into long range triplet correlations (LRTCs) with $S_z=1$. Unlike singlet correlations and SRTCs, LRTCs are not as susceptible to depairing and therefore can maintain coherence over relatively large distances across a FM. The most common method to experimentally investigate spin-polarized supercurrent is via transport experiments in Josephson junctions (JJs) in which the weak link contains magnetic non-collinearities, for example, non-collinear ferromagnetic layers,\cite{Khaire_2010, Gingrich2012Dec} helical ferromagnets,\cite{Robinson_2010} Heusler alloys\cite{Sprungmann_2010} and artificial domain walls.\cite{Robinson2012Sep} Most recently, long-range supercurrents have been observed in JJs with a chiral Kagome antiferromagnet as the weak link.\cite{Jeon_2021} Historically, LRTCs have been demonstrated in JJs with half-metallic ferromagnetic oxides, some with collinear magnetic structures.\cite{Keizer_2006, Dybko_2009, Anwar_2010} These systems host LRTCs over significantly longer distances than conventional FMs. However, the transport through these JJs is more complicated as the half-metallic weak link also acts as a spin filter.\cite{Eschrig2003Apr, Eschrig2008} 

As an alternative to a magnetic non-collinearity, spin-orbit coupling (SOC) at the SC/FM interface has been proposed as a source of LRTCs.\cite{Niu_2012, Bergeret_2014, Konschelle_2014, Jacobsen2016Apr,Amundsen2022Oct} The required SOC at the interface can be obtained by placing a thin heavy metal layer between the SC and the FM. The optimal magnetization direction of the FM depends on the geometry of the JJ: the FM should be canted at $45\degree$ from the direction of the supercurrent in a vertical JJ\cite{Jacobsen2016Apr, Amundsen2019Aug} or directed along the supercurrent for a lateral JJ.\cite{Eskilt2019Dec} Despite the reduced complexity of magnetic layering, earlier transport studies in JJs haven't established the presence of LRTCs in these systems.\cite{Satchell_2018, Satchell_2019, Satchell2021May} However, recent studies with non-equilibrium spin-pumping experiments have established the connection between spin absorption and LRTCs in the presence of SOC.\cite{Jeon2018Jun, Jeon2019Jan, Chan2023Oct}

In this work, we investigate transport in vertical JJs with the structure SC/NM/FM/NM/SC where NM are non-magnetic layers formed from Pt, Ir, Cu or Pt/Cu and FM are ferromagnetic layers with perpendicular magnetic anisotropy (PMA) formed from Co/Ni/Co multilayers. For the case of Pt/Cu, Pt is in each case, sandwiched between the SC layer and Cu. We vary the thickness of the FM for a given NM layer thickness. We show that the critical current ($I_{\text{c}}$) in JJs with Pt and Ir layers decays similarly with FM thickness. On the other hand, this decay in `Pt' and `Ir' JJs is qualitatively different from that in JJs with `Cu' and `Pt/Cu'. We compare these dependencies with the established $0-\pi$ mechanism and conclude the possible presence of LRTCs in our `Pt' and `Ir' JJs.

Multi-layer thin films with the structure of Si/SiO$_x$/ TaN(2)/Nb(50)/NM/Co(0.7)/[Ni(0.3)/Co(0.7)]$_n$/NM/ Nb(10)/Pt(2) were grown on Si(100) in a high vacuum sputter deposition system, where $n$ is the number of Ni/Co bilayer repeats (see supplementary material Sec. I for more details). Note that the layer thicknesses are given in nanometers throughout the text unless otherwise noted. A few `Pt' samples were grown on a thinner Nb(20) layer. In the stack, the top Nb(10) layer ensures that the SC/NM and NM/SC interfaces in the junctions are as similar as possible and the Pt capping layer helps to prevent oxidation. Atomic force microscopy (AFM) measurements of the films show smooth growth with rms roughnesses averaging between \SI{0.14}{} and \SI{0.22}{\nano\meter} over an area of $\SI{1}\times\SI{1}{\micro\meter\squared}$. The superconducting critical temperature ($T_c$) of the bottom Nb layer, as determined from transport measurements in the complete stack, is ~\SI{7.5}{\kelvin} for Nb(50) and \SI{5.9}{\kelvin} for Nb(20) in the absence of any external magnetic field. Vibrating sample magnetometry (VSM) of the films was performed at room temperature using a LakeShore 8600 Series magnetometer. Magnetization mapping at room temperature was carried out using polar magneto-optical microscopy (p-MOKE).

\begin{figure}[!hbtp]
     \centering
     \begin{minipage}{0.23\textwidth}
     \begin{subfigure}[b]{\textwidth}
         \centering
         \includegraphics[width=\textwidth]{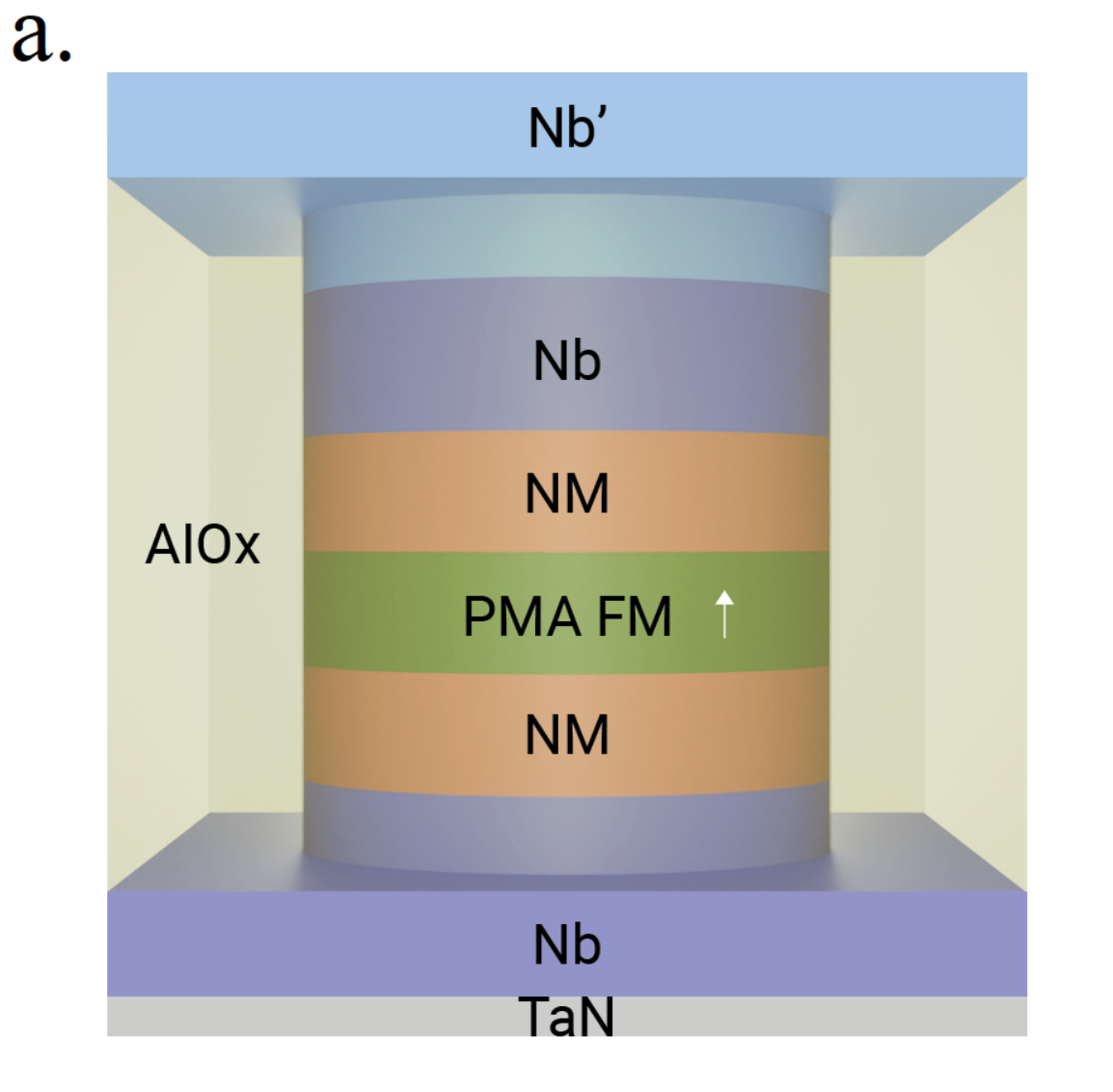}
         \phantomsubcaption
         \label{fig:JJStack}
     \end{subfigure}
     \begin{subfigure}[b]{\textwidth}
         \centering
         \includegraphics[width=\textwidth]{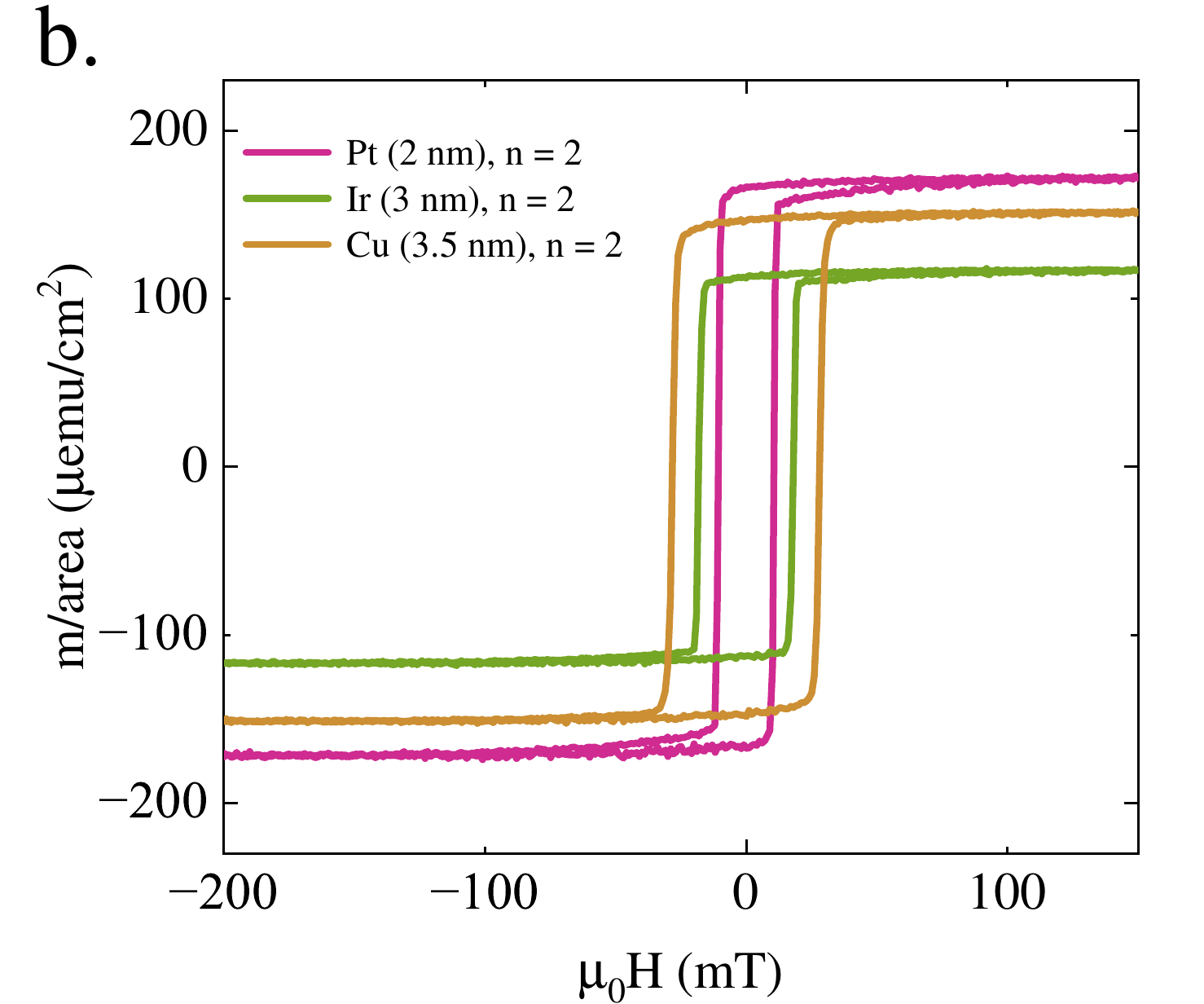}
         \phantomsubcaption
         \label{fig:MvH}
     \end{subfigure}  
     \end{minipage}
     \begin{minipage}{0.23\textwidth}
     \begin{subfigure}[b]{\textwidth}
         \centering
         \includegraphics[width=\textwidth]{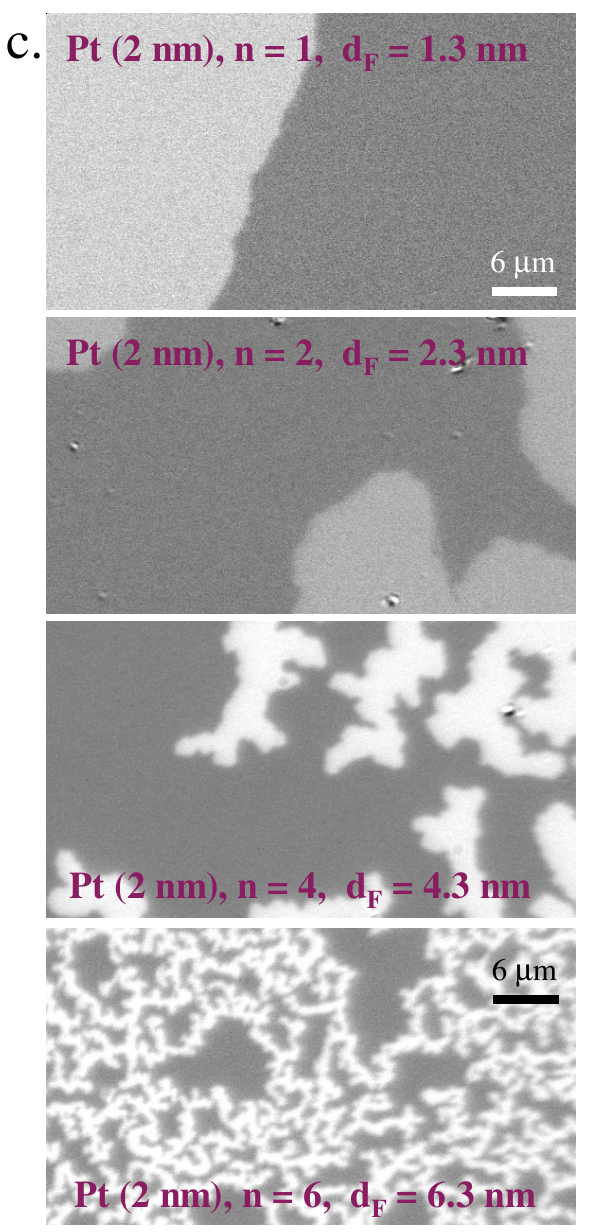}
         \phantomsubcaption
         \label{fig:MOKE}
     \end{subfigure}
     \end{minipage}
        \caption{(a) Schematic structure of the JJs investigated in this work. (b) Dependence of OOP magnetic moment/area ($m$/area) as a function of OOP magnetic field ($H$) of films with Pt, Cu and Ir layers, where $n = 2$. (c) p-MOKE of films with Pt(2) layers for different FM layer thicknesses ($d_{\text{F}}$).}
        \label{fig:Magnetic}
\end{figure}

The films were subsequently patterned into JJs via photo-lithography and Ar-ion milling etching processes (see supplementary material Sec. I for fabrication details). The structure of a fabricated JJ is schematically depicted in Fig. \ref{fig:JJStack}. The transport measurements were carried out in a Quantum Design PPMS using Keithley 2182a nanovoltmeters and 6221 current sources.


Fig. \ref{fig:MvH} shows the out-of-plane (OOP) moment per unit area as a function of OOP magnetic field for typical films before device fabrication. All films exhibit PMA and have nearly 100\% remnant magnetization. We note an additional magnetization for films with Pt layers compared to those with Cu and Pt/Cu, that arises from the proximity-induced magnetization in Pt.\cite{Geissler_2001} Conversely, films with Ir layers exhibit the lowest magnetization although Ir exhibits a relatively smaller proximity-induced magnetization.\cite{Ryu_2013} In Fig. \ref{fig:MOKE} we compare the magnetization switching process of films with NM = Pt(\SI{2}{\nano\meter}) and $n$ = 1, 4 and 6 (which correspond to summed FM layer thicknesses, $d_{\text{F}}$ = \SI{1.3}{}, \SI{4.3}{} and \SI{6.3}{\nano\meter}). The increase of $d_{\text{F}}$ results in a denser domain structure as it is energetically favorable. Thus, we observe a gradual transition in the magnetization switching mechanism from the propagation of a single domain wall (DW) to the percolation of stripe domains.

\begin{figure}[!htbp]
    \begin{subfigure}[b]{\textwidth}
        \phantomsubcaption
        \label{fig:Frh1}
    \end{subfigure}
    \begin{subfigure}[b]{\textwidth}
         \phantomsubcaption
         \label{fig:Frh2}
    \end{subfigure}
    \begin{subfigure}[b]{\textwidth}
         \phantomsubcaption
         \label{fig:Frh3}
    \end{subfigure}
\includegraphics[width=\linewidth]{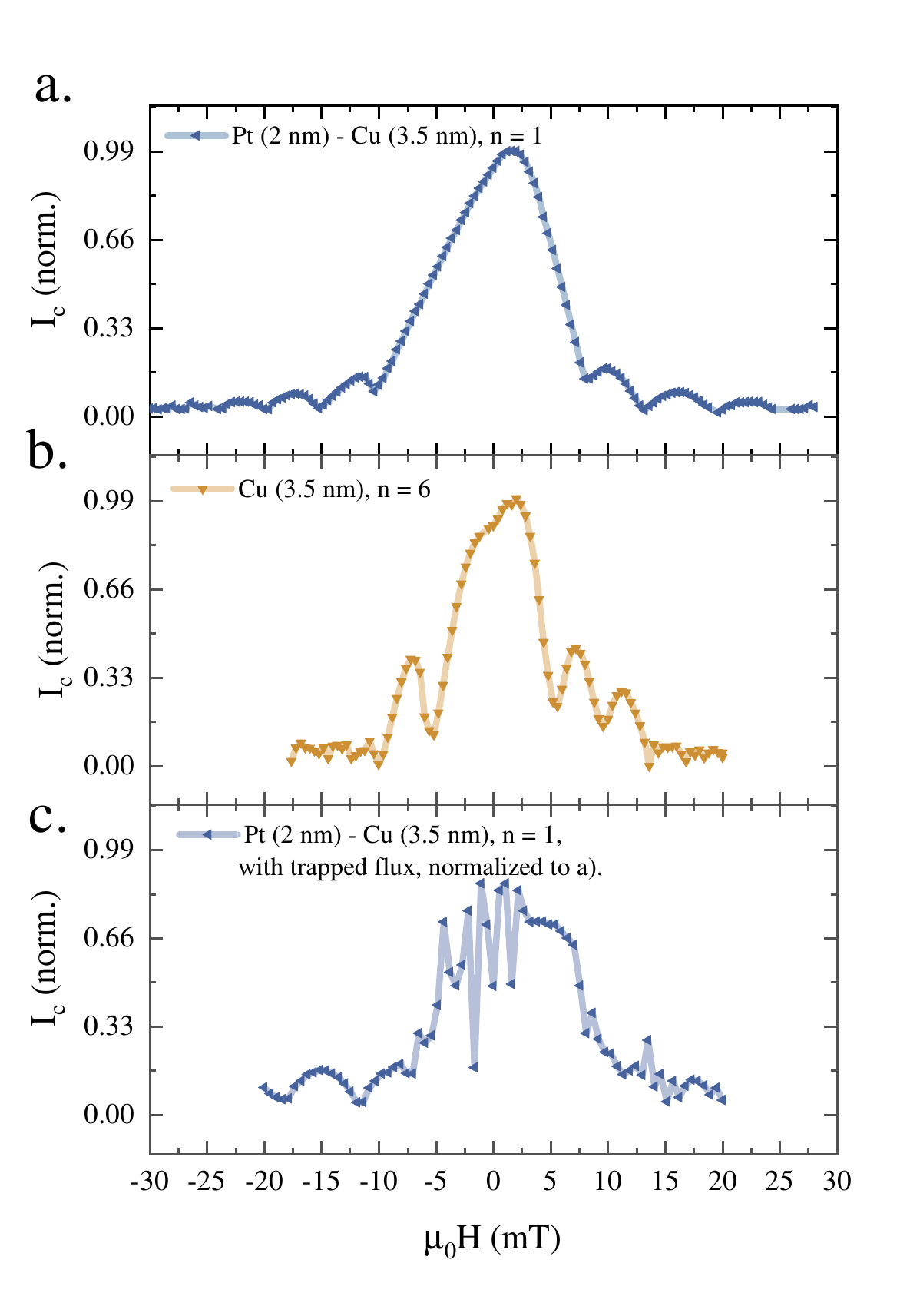}
\caption{Normalized critical current ($I_c$) as a function of external in-plane magnetic field ($H$) showing magnetic diffraction of JJs for three different cases: (a) $n=1$ and NM = Pt/Cu, (b) $n=6$ and NM = Cu, (c) $n=1$ and NM = Pt/Cu with trapped flux, normalized to (a).}
\label{fig:Fraunhofers}
\centering
\end{figure}

Fig. \ref{fig:Fraunhofers} shows various cases of the dependence of the normalized critical current ($I_c$) on the external in-plane magnetic field, capturing the magnetic diffraction pattern measured in our junctions, where $I_c$ was extracted from fitting the voltage ($V$) versus current ($I$) curves to the relation, $V = \operatorname{Re}(R_N \sqrt{I^2-I^2_c})$, where $R_N$ is the resistance of the junction in the dissipative state.\cite{Barone1982Jul} Fig. \ref{fig:Frh1}  represents a typical diffraction pattern for the `Cu' and `Pt/Cu' JJs with the thinnest FM layer i.e. $n=1$. A skewed pattern indicates self-field effects arising from the magnetic field generated by the large supercurrent in these JJs. However, the self-field effect is not expected to considerably alter the maximum critical current in the diffraction pattern.\cite{Golod2022Jun} We do not observe skewing for any of the `Pt' and `Ir' JJs or for `Cu' and `Pt/Cu' JJs with $n\geq2$ due to reduced Josephson coupling and lower supercurrents, and we measure an expected Airy diffraction pattern with minor deviations at higher order diffraction lobes. Fig. \ref{fig:Frh2} represents a typical diffraction pattern for junctions with thicker magnetic layers measured for `Cu' JJs with $n=6$. With increasing $n$, we observe increasing deviations from Fig. \ref{fig:Frh1}. These patterns are asymmetrical with respect to zero field and the local maxima of the lobes decrease non-monotonically. Such diffraction patterns are often measured in JJs formed from Co/Ni multilayers with relatively large diameters.\cite{Gingrich2012Dec,Satchell_2018,Satchell_2019} This has been associated with the spatial non-uniformity of the magnetic field inside a JJ .\cite{Golovchanskiy2016Dec, Krasnov2020Apr}  We find that the diffraction pattern is highly dependent on the magnetic initialization procedure of our junctions before cooling to below $T_c$. Saturating the PMA layer by applying an OOP $H=\SI{1}{\tesla}$ before cooling results in a wide and prominent peak centered around zero field. However, fitting to an Airy-like function leads to errors due to asymmetry in the pattern. Therefore, we obtain the $I_{\text{c}}$ maxima of the JJ from the experimental data without fitting. In Fig. \ref{fig:Frh3} we show the diffraction measured on the same device as Fig. \ref{fig:Frh1} but under different cooling conditions, highlighting the unwanted effects of trapped flux within the junction. Unlike the continuous patterns observed in Fig. \ref{fig:Frh1} and Fig. \ref{fig:Frh2}, Fig. \ref{fig:Frh3} exhibits clear discontinuities which have been attributed to trapped flux in the JJs.\cite{Golod2010Jun} The issue of Abrikosov vortices (AVs) is significant as it complicates accurate identification of the maximum $I_{\text{c}}$. To mitigate parasitic effects we perform the following protocol until the desired condition without AVs is achieved: we measure diffraction pattern at the working temperature, \SI{2}{\kelvin}; if discontinuities are detected, we warm the device to \SI{10}{\kelvin}, saturate it with an OOP $H=\SI{1}{\tesla}$, cool slowly to the working temperature, and then re-examine the pattern.

\begin{figure*}
    \begin{subfigure}[b]{\textwidth}
        \phantomsubcaption
        \label{fig:ICRN1}
    \end{subfigure}
    \begin{subfigure}[b]{\textwidth}
         \phantomsubcaption
         \label{fig:ICRN2}
    \end{subfigure}
    \begin{subfigure}[b]{\textwidth}
         \phantomsubcaption
         \label{fig:ICRN3}
    \end{subfigure}
    \begin{subfigure}[b]{\textwidth}
         \phantomsubcaption
         \label{fig:ICRN4}
    \end{subfigure}
    \includegraphics[width=\textwidth]{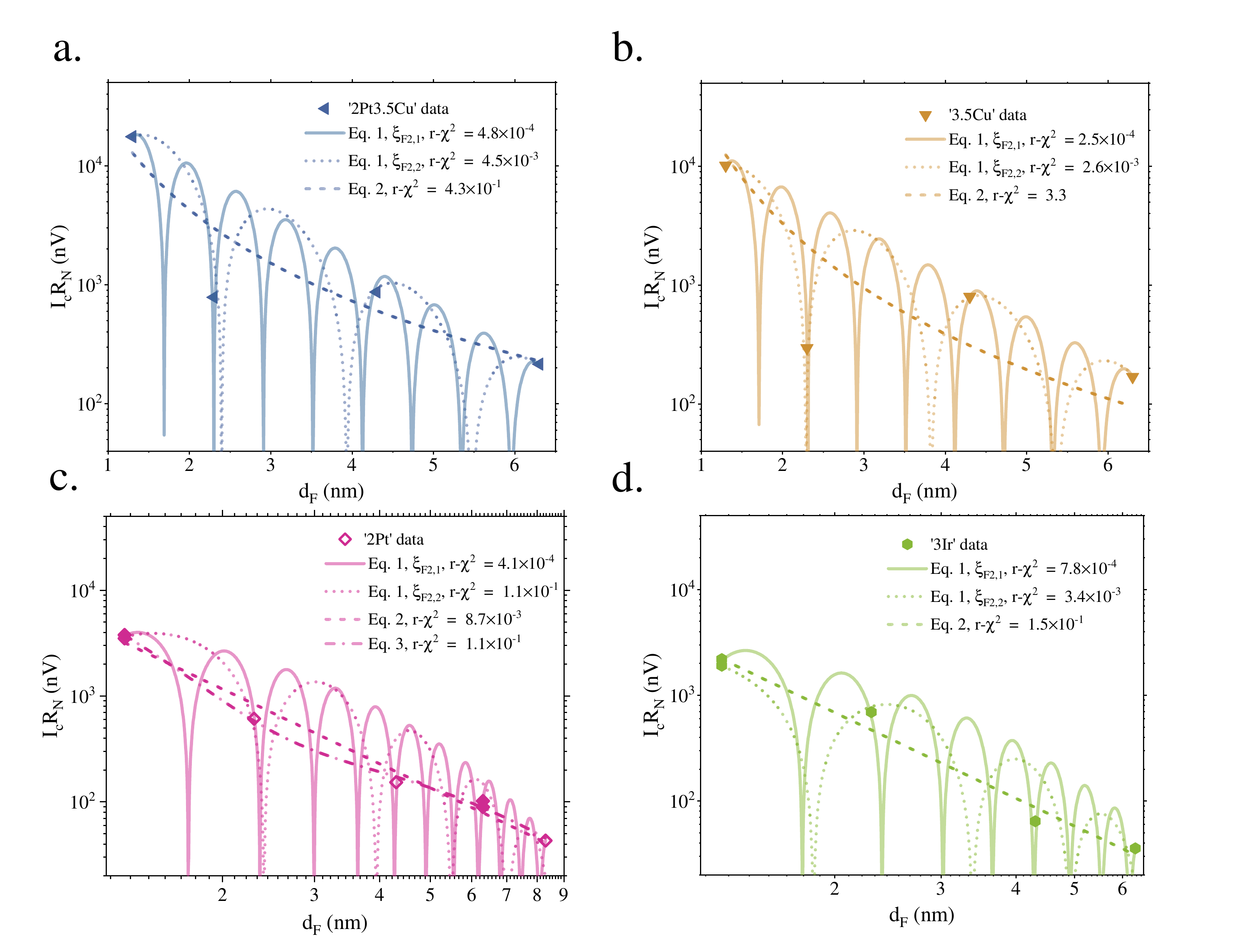}
    \caption{$I_{\text{c}} R_{\text{N}}$ values of JJs with fixed NM layers and varying thickness of FM, $d_F$, where FM is Co(0.7)/[Ni(0.3)/Co(0.7)]$_n$, $r-\chi^2$ is reduced $\chi^2$ values of the respective fits. (a) `2Pt3.5Cu' JJs with Pt(2)/Cu(3.5)/FM/Cu(3.5)/Pt(2) weak links in semi-log scale, (b) `3.5Cu' JJs with Cu(3.5)/FM/Cu(3.5) weak links in semi-log scale (c) `2Pt'JJs with Pt(2)/FM/Pt(2) weak links in log-log scale, solid magenta points correspond to `2Pt' JJs with Nb(20) at the bottom, and hollow points correspond to the `2Pt' JJs with Nb(50). (d) `3Ir' JJs with Ir(3)/FM/Ir(3) weak links in log-log scale.}  
    \label{fig:ICRNvsFM}
\end{figure*}

In Fig. \ref{fig:ICRNvsFM}, we plot the decay of $I_{\text{c}}R_{\text{N}}$ across JJs with fixed NM layers (referred to as `3.5Cu', `2Pt', `3Ir' and `2Pt3.5Cu', where the numbers preceding the layers are their thicknesses in \SI{}{\nano\meter}) for varying $n$.  We fix Pt thickness at \SI{2}{\nano\meter} and Ir thickness at \SI{3}{\nano\meter} for further experiments  because at these thicknesses we obtain an optimum balance between superconducting properties, magnetic properties and flux channeling (see supplementary Sec. II). $I_{\text{c}}$ is multiplied by $R_{\text{N}}$, to account for variations in JJ area during device fabrication. However, this does not account for variations in $I_{\text{c}}$ due to supercurrent transmission across different layers and their interfaces. The highest $I_{\text{c}} R_{\text{N}}$ values at $n = 1$ are observed for `2Pt3.5Cu' and `3.5Cu' JJs, while those for `2Pt' and `3Ir' JJs are nearly an order of magnitude lower. This disparity may be attributed to the difference in supercurrent transmission across the Cu/Co, Pt/Co and Ir/Co interfaces. Reduced Josephson coupling in JJs with Pt/Co and Ir/Co might be connected to the strong interfacial Dzyaloshinskii-Moriya exchange interaction at these interfaces\cite{Ryu_2013} which may lead to additional depairing mechanisms. The Nb/Cu and Nb/Pt interfaces appear to contribute similarly to the attenuation of $I_{\text{c}} R_{\text{N}}$, since the values for the `2Pt3.5Cu' and `3.5Cu' layers are nearly the same. We also highlight the difference in supercurrent transmission between Cu/Co and Cu/Ni interfaces in the supplementary material (see supplementary Sec. IV), noting that Cu/Ni has higher $I_{\text{c}} R_{\text{N}}$. The variability of attenuation at interfaces with different materials complicates the identification of LRTC in JJs with SOC layers based solely on the absolute values of $I_{\text{c}} R_{\text{N}}$.

Turning to Fig. \ref{fig:ICRNvsFM}, the data can be fitted with the established model of $0-\pi$ oscillations in ferromagnetic JJs:\cite{Buzdin1991}

\begin{equation}\label{Eq.1}
    I_c R_N (d_F) = I_{c, 0} R_N \exp\left( \frac{-d_F}{\xi_{F1}} \right) \left| \sin\left(\frac{d_F-d_{0-\pi}}{\xi_{F2}}\right) \right|
\end{equation}
where $I_{c,0}$ is an effective supercurrent at $d_F=$ \SI{0}{\nano\meter} and $R_N$ is the resistance of the junction in the normal state, $d_{0-\pi}$ is $d_F$ corresponding to the first $0-\pi$ oscillation, $\xi_{F1}$ is a decay constant in the FM, and $\xi_{F2}$ is the electron coherence length in FM. For dirty FM weak links, the diffusive regime is used where $\xi_{F2} = \sqrt{\hbar D_F /E_{\text{ex}}}$ in Eq. \ref{Eq.1}. Here, $D_F = v_F l_{\text{mfp}}/3 $ is the diffusion constant, $v_F$ is the Fermi velocity, $l_{\text{mfp}}$ is the electron mean free path and $E_{\text{ex}}$ is the exchange energy of the FM. In case of strong FM such as Co and Ni with $E_{ex}\gg \Delta$, intermediate limit is typically considered, where  $\xi_{F2}$ is  ballistic in Eq. \ref{Eq.1} with $l_{\text{mfp}} \gg \xi_{F2} = \hbar v_{F}/2E_{ex}$. \cite{Bergeret_2001} Large number of alternating sub-nanometer Co and Ni layers may lead to an additional scattering of Cooper pairs resulting in reduction of $l_{\text{mfp}}$, affecting the intermediate limit  condition.\cite{Bass2016Jun} We note that for JJs with Co/Ni/Co multilayers, only $\xi_{F1}$ has been reported,\cite{Gingrich2012Dec} while $0-\pi$ oscillations and $\xi_{F2}$ have not yet been established. The main reason for this is that $\xi_{F2}$ is comparable to the thickness of a single Co/Ni repeat, making it challenging to resolve the oscillations. Additionally, due to the differing exchange energies of Ni and Co, the frequency of the oscillations is spatially modulated, which may also lead to additional  deviations from the model. In our case, we assume that the FM has an average uniform exchange energy across the weak link and compare both ballistic and diffusive $\xi_{F2}$. We consider several factors to check the physical feasibility of such fits. First, $\xi_{F2}$ should be similar in all four types of junctions, as we use identical FMs with identical $E_{\text{ex}}$. In Fig. \ref{fig:ICRN2} and Fig. \ref{fig:ICRN1}, corresponding to `3.5Cu' and `2Pt3.5Cu', we observe that the data exhibits a prominent dip which we associate with an oscillation. Typically, in strong ferromagnets $d_{0-\pi},  \xi_{F2} < \xi_{F1}$.\cite{Birge2024Jan} To fulfill this condition, the dip should not be less than the first $\pi-0$ transition.  At the same time, the dip cannot be higher than second $0-\pi$ transition, as such fit would produce an unrealistically high $E_{\text{ex}}$. This leads us to two possible values for all data sets: $\xi_{F2,1} =$  \SI{0.2}{\nano\meter} and $\xi_{F2,2} = \SI{0.5}{\nano\meter}$. In the intermediate limit, \cite{Bergeret_2001} we calculate exchange energy as  $E^i_{\text{ex}} = \hbar v_{F}/2\xi_{F2}$. Assuming $v_F \approx 0.3 \times 10^6$ m/s,\cite{Robinson_2010} we get $E^i_{\text{ex},1} = 0.55$ eV for $\xi_{F2,1} = 0.2 $ nm and  $E^i_{\text{ex},2} = 0.22$ eV for $\xi_{F2,2} = 0.5 $ nm. On the other hand, in fully diffusive limit, $E^d_{\text{ex}} = \hbar D_{F} / \xi_{F2}^2 \approx \hbar v_F \xi_{F1} / 3 \xi_{F2}^2$, where we estimated $l_{\text{mfp}}\approx \xi_{F1}=$ \SI{1.15}{\nano\meter}. Here, $\xi_{F2,1} =$ \SI{0.2}{\nano\meter} gives $E^d_{\text{ex},1} = 1.9$ eV and $\xi_{F2,2} = $ \SI{0.5}{\nano\meter} corresponds to $E^d_{\text{ex},2}=0.3$ eV. $E^d_{\text{ex},1}$ is significantly higher than the values typically reported for Co or Ni, and $E^i_{\text{ex},1}$, $E^i_{\text{ex},2}$, $E^d_{\text{ex},2}$ are realistic values that one would expect .\cite{Robinson2009Nov, Robinson_2010} Finally, we expect $d_{0-\pi}$ to be inversely related to the magnetic moments of the films in Fig. \ref{fig:MvH} due to proximity-induced magnetization effects,  $d_{0-\pi, \text{`2Pt'}}< d_{0-\pi, \text{`3.5Cu'}}< d_{0-\pi,\text{ `3Ir'}}$. This condition can only be satisfied for the case of $\xi_{F2,1} =$ \SI{0.2}{\nano\meter} and $d_{0-\pi,\text{ `3Ir'}} = $ \SI{0.51}{\nano\meter}, $d_{0-\pi,\text{ `3.5Cu'}} = $ $\SI{0.5}{\nano\meter}$ and $d_{0-\pi,\text{ `2Pt'}} =$ $\SI{0.45}{\nano\meter}$. Therefore, we obtain physically feasible fit to the $0-\pi$ model only in the case of the intermediate limit and $\xi_{F2} = 0.2 $ nm. Additionally, we observe linear dependence of $I_c$ on temperature in both `3.5Cu' and `2Pt' JJs for $d_F =$ \SI{6.3}{\nano\meter} (see supplementary material Sec. VI). This confirms validity of the intermediate regime, as a fully diffusive case would be expected to show a superlinear dependence.\cite{Kapran2021Mar} We also note that this analysis indicates reduced decay for ‘2Pt’ JJs, which has about $20\%$ higher $\xi_{F1}$ than all other junctions, while ‘3Ir’ JJs are comparable to `3.5Cu' and `2Pt3.5Cu' (see supplementary material Sec. III). 

To fit `2Pt' and `3Ir' data to $0-\pi$ model we need to assume that  all data points are located near the $0-\pi$ transitions. Alternatively, if we now suppose that the data points for ‘2Pt’ and ‘3Ir’ are outside of dips and focus on the envelopes of the oscillating components, we find a good fit to an inverse power law dependence of the $I_cR_N$ on $d_F$ as follows:

\begin{equation}\label{Eq.2}
    I_c R_N (d_F) \propto d_F^{-\alpha}
\end{equation}
where $\alpha_{\text{‘2Pt’}} = 2.39$ and $\alpha_{\text{‘3Ir’}} = 2.75$. The algebraic dependence in Eq. \ref{Eq.2} could indicate an enhancement of the mean free path and reduced scattering of Cooper pairs in the magnetic material, suggesting that transport in the JJs is more ballistic than in the intermediate regime. If we focus only on the exponential envelopes, we can assume a crossover of two contributions — short-ranged (SR) and long-ranged (LR) — where the former is dominant for smaller $d_F$:

\begin{equation}\label{Eq.3}
    \begin{aligned}
    I_c R_N (d_F) \propto  I_{c,0}^{\text{SR}}R_N \exp \left( \frac{-d_{F}}{\xi_{F1}^{\text{SR}}} \right) + \\
    I_{c,0}^{\text{LR}}R_N \exp \left(\frac{-d_{F}}{\xi_{F1}^{\text{LR}}}\right), I_{c,0}^{\text{SR}}R_N \gg I_{c,0}^{\text{LR}}R_N 
    \end{aligned}
\end{equation}

Both these models (Eq. \ref{Eq.2} and \ref{Eq.3}) imply the presence of LRTCs. In the latter case, fitting ‘2Pt’ data to Eq. \ref{Eq.3} produces $\xi^{\text{LR}}$ = \SI{3}{\nano\meter} and $\xi^{\text{SR}}$ = \SI{0.4}{\nano\meter}. Although $\xi^{\text{SR}}$ = \SI{0.4}{\nano\meter} is much shorter compared to ‘3.5Cu’ and ‘2Pt3.5Cu’ JJs from previously considered models, the ratio $\xi^{\text{LR}}/\xi^{\text{SR}}=7.5$ is a realistic reduction of decay from LRTC with respect to SRTC in the studied FM (see supplementary material Sec. IV). We note that theory predicts that the LRTC generation in our system depends on the relative orientation of the magnetization with respect to the direction of current flow, peaking at $45\degree$, and vanishing at $0\degree$ or $90\degree$.\cite{Jacobsen2016Apr,Amundsen2019Aug} Therefore, LRTCs are unlikely in JJs with fully OOP saturated FM. The dependence of supercurrent on the angle of magnetization naturally brings into consideration the magnetic origin of Eq. \ref{Eq.2}. One possible way to reliably achieve $45\degree$ magnetization is by utilizing a domain wall (DW). Indeed, the magnetization in a $180\degree$ DW in a FM with PMA will always contain $45\degree$ and $-45\degree$  orientations, regardless of the thickness of the FM or the lateral size of the magnetic film. Next, junctions with thicker $d_F$ are expected to have a higher density of DWs as was demonstrated in Fig. \ref{fig:MOKE}. The asymmetric diffraction pattern in Fig. \ref{fig:Frh2} resulting from non-uniform magnetic field in the weak link, is consistent with this hypothesis. However, this would also imply that the Meissner effect takes part in a competition between exchange, anisotropy and magnetostatic energies in the FM and that DWs are spontaneously nucleated at the superconducting transition of Nb. In additional $M(T)$  measurements on an array of \SI{6}{\micro\meter} pillars for `3.5Cu' S/FM(n=6)/S (see supplementary material Sec. V), we do not find any indication of such nucleation at $T_c$. Therefore, we can rule out spontaneous DW nucleation. On the other hand, the absence of DWs does not exclude possible canting because of the Meissner effect in Nb. By ensuring the absence of AVs in the Nb of the JJs via our initialization protocol, we expect the SC layer to expel the field lines into the NM layers between the SC and FM layers, directing the lines tangentially to the surface of the FM. This channeling of the field may act as an effective external in-plane field $H_{IP}^{\text{Meiss}}\propto M_s \propto d_F$, and thereby introduce a slight canting of the FM moment without breaking into DWs. Therefore, the magnetostatic interaction of the PMA FM with SC may lead to a reduction of supercurrent decay in JJs with SOC layers.

In conclusion, we have investigated vertical FM JJs with Pt, Ir, Cu and Pt/Cu layers sandwiched between a FM and the SC. The $I_cR_N$ versus $d_F$ data for JJs with Pt/Co and Ir/Co interfaces can be fitted with the established $0-\pi$ model with the assumption that the data is aliased by $0-\pi$ oscillations. From this analysis, we observe that Josephson coupling decays slower in JJs with Pt/Co interfaces with increasing $d_F$ than in JJs with Cu/Co interfaces. We also suggest an inverse power law dependence of $I_cR_N$ on $d_F$ in JJs with Pt/Co and Ir/Co and speculate its origin being a result of the interaction between the magnetization and the Meissner effect. Although the supercurrent transmission is attenuated, its reduced decay with $d_F$ may indicate an unoptimized generation of $S_z=1$ spin-triplet Cooper pairing in JJs with the SOC layers. We note that the choice of materials in this study makes our JJs compatible with contemporary spintronic devices.\cite{Ryu_2013}


\section*{Supplementary Material}
See supplementary material for details concerning sample preparation and fabrication, a comparison of different SOC thicknesses, a table with parameters from fitting models, a comparison of different FM types, a verification of the absence of DWs and the dependence of JJ critical current on temperature.

\begin{acknowledgments}
We thank J-C. Jeon, J. Yoon, Y. Guan, P.K. Sivakumar for valuable discussions. We also thank P. Grunewald, H. Deniz, D. Knyazev and A. Stakhnova for technical help. We thank the reviewers for their feedback as it resulted in a significantly improved manuscript.\\

Funded by the European  Union (ERC Advanced Grant SUPERMINT, project number  101054860). Views and opinions expressed are however those of the authors only and do not necessarily reflect those of the European Union or the European Research Council. Neither the European Union nor the granting authority can be held responsible for them.
\end{acknowledgments}

\section*{Author Declarations}
\subsection*{Conflict of interest}
The authors have no conflicts to disclose.

\section*{Data Availability}
The supporting data are available from the corresponding author upon reasonable request.

\bibliography{references}

\end{document}


\title{Supplementary Material: Reduced decay in Josephson coupling across ferromagnetic junctions with spin-orbit coupling layers}


\maketitle

\section{Sample preparation and device fabrication details}
The films were grown on Si(100) wafers with \SI{300}{\nano\meter}-thick thermally oxidized $\text{SiO}_2$ layers in an ultra-high vacuum multi-source sputter deposition system, which has a typical base pressure below \SI{e-7}{\pascal}. The TaN seed layer was deposited by reactive sputtering using $\text{Ar/N}_2$ sputter gas. The Nb layers were grown by ion beam sputter deposition at a pressure of \SI{9e-3}{\pascal}. All the remaining layers were grown by dc magnetron sputtering at an Ar pressure of \SI{0.4}{\pascal}. All the films were grown at ambient temperature.

The complete stack was patterned and etched to form the bottom Nb electrodes. Then, circular pillars with diameters of \SI{4}{} and \SI{6}{\micro\meter} were patterned and etched down to the bottom NM layer using an in-situ secondary ion mass spectrometer (SIMS) end point detection. Without breaking vacuum, a \SI{60}{\nano\meter} thick layer of insulating $\text{AlO}_x$ was deposited in-situ to electrically isolate the bottom SC leads to avoid shorting. Lastly, the top electrodes were made using photo-lithography followed by the deposition of Nb(50)/Au(4) layers grown using ion beam sputter deposition at a pressure of \SI{1.2e-2}{\pascal} after light Ar-ion etching to completely remove the Pt capping layer in the pillars. The top Nb layer has a $T_c$ of $\sim \SI{7.2}{\kelvin}$.

\section{$I_cR_N$ for varying thickness of SOC layers and fixed $d_F$ at $n=1$}

\begin{figure}[htbp]
    \includegraphics[width=0.5\textwidth]{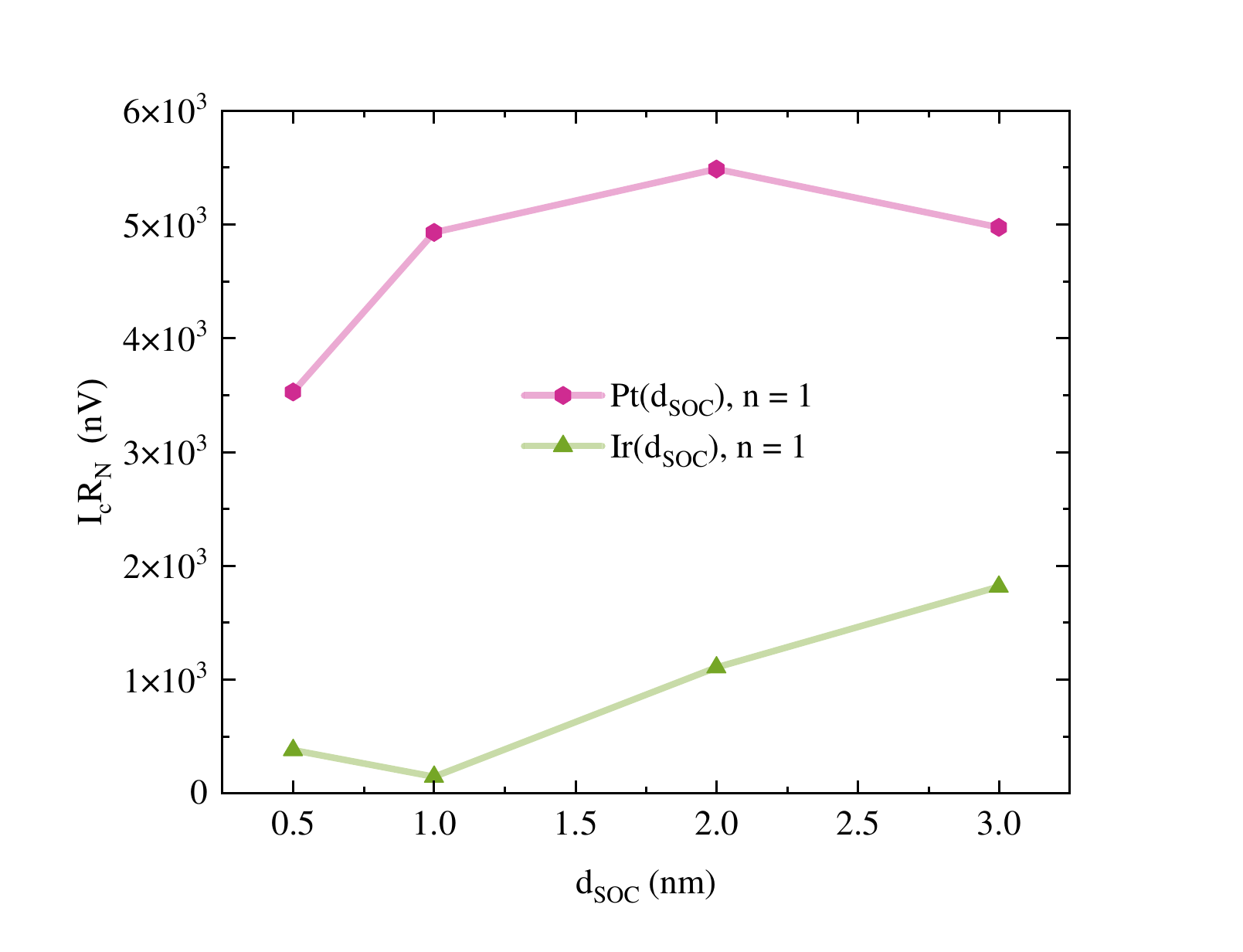}
    \caption{$I_{\text{c}} R_{\text{N}}$ values of JJs with varying thickness of Pt and Ir layers ($d_{\text{SOC}}$) and fixed FM thickness, $n=1$ ($d_{\text{F}} = \SI{1.3}{\nano\meter}$)}
    \label{fig:IcRNvsSOC}
\end{figure}

In Fig. \ref{fig:IcRNvsSOC}, we vary the thickness of Pt and Ir layers with strong SOC ($d_{\text{SOC}}$), while fixing $n=1$ ($d_{\text{F}} = \SI{1.3}{\nano\meter}$), and measure $I_{\text{c}} R_{\text{N}}$ values. We note that these samples belong to a different deposition run and were processed in a different batch than the JJs in the main text. Although the $I_c$s for these are comparable to the JJs in the main text, the $R_N$ is about 20\% higher, thus comparably inflating the $I_c R_N$ values somewhat. Therefore, we do not include this data for '2Pt', n = 1 JJs in Fig.3 of main text. We observe a non-monotonic dependence of $I_{\text{c}} R_{\text{N}}$, with maximum values at NM = Pt(2) and Ir(2). For films with the thinnest SOC layers, NM = Pt(0.5) and Ir(0.5), the remanent OOP magnetization is zero, while films with $d_{\text{SOC}}\geq$ \SI{1}{\nano\meter} exhibit close to 100\% remanent magnetization. At the same time, we find that proximity-induced magnetization effects are saturating in our films at $d_{\text{SOC}}\approx$  \SI{1}{\nano\meter}. On the other hand, our measurements indicate that flux trapping is more prominent for the JJs with the thinner SOC layers, $d_{\text{SOC}}\leq$ \SI{2}{\nano\meter} for Pt and $d_{\text{SOC}}\leq$ \SI{3}{\nano\meter} for Ir. AVs in the Nb electrodes are expected to be spontaneously generated at the edge of an SC/FM interface with SOC when the magnetization is directed orthogonal to the edge.\cite{OldeOlthof2019Dec} However, we attribute this not only to the aforementioned effect but also to the field focusing effect in the non-magnetic layers due to the Meissner expulsion of the field.\cite{Golod2019Nov} We find that the flux trapping in the Nb electrodes is reduced by including an additional layer, such as Cu,  between the SOC and FM layers. We fix Pt thickness at \SI{2}{\nano\meter} and Ir thickness at \SI{3}{\nano\meter} for further experiments with $n$ variation because at these thicknesses we obtain an optimum balance between superconducting properties, consistency of magnetic properties and flux channeling.


\section{Parameters obtained from fitting in the main text}


In Table \ref{tab:pars} we provide fitting parameters from all equations used in the main text.

\begin{table}[htbp]
\centering
\begin{tabular}{|c|c|c|c|c|}
\hline
\textbf{} & \textbf{`2Pt'} & \textbf{`3Ir'} & \textbf{`3.5Cu'} & \textbf{`2Pt3.5Cu'} \\ \hline
        $\xi_{F1,1}$ & \SI{1.6}{\nano\meter} & \SI{1.3}{\nano\meter} & \SI{1.3}{\nano\meter} & \SI{1.1}{\nano\meter} \\ \hline
        $\xi_{F1,2}$ & \SI{1.4}{\nano\meter} & \SI{1.1}{\nano\meter} & \SI{1.2}{\nano\meter} & \SI{1.1}{\nano\meter} \\ \hline
        $\xi_{F2,1}$ & \SI{0.2}{\nano\meter} & \SI{0.2}{\nano\meter} & \SI{0.2}{\nano\meter} & \SI{0.2}{\nano\meter} \\ \hline
        $\xi_{F2,2}$ & \SI{0.5}{\nano\meter} & \SI{0.5}{\nano\meter} & \SI{0.5}{\nano\meter} & \SI{0.5}{\nano\meter} \\ \hline
        $I_{c,0,1}R_N$ & \SI{9.6}{\micro\volt} & \SI{8.1}{\micro\volt} & \SI{30.9}{\micro\volt} & \SI{62.5}{\micro\volt} \\ \hline
        $I_{c,0,2}R_N$ & \SI{12.5}{\micro\volt} & \SI{13.5}{\micro\volt} & \SI{32.6}{\micro\volt} & \SI{74.4}{\micro\volt} \\ \hline
        $d_{0-\pi,1}$ & \SI{0.45}{\nano\meter} & \SI{0.51}{\nano\meter} & \SI{0.5}{\nano\meter} & \SI{0.47}{\nano\meter} \\ \hline
        $d_{0-\pi,2}$ & \SI{0.91}{\nano\meter} & \SI{1.04}{\nano\meter} & \SI{0.75}{\nano\meter} & \SI{0.86}{\nano\meter} \\ \hline
        $\alpha$ & -2.3 & -2.7 & -3.1 & -2.55 \\ \hline
        $\xi_{F1}^{\text{SR}}$ & \SI{0.4}{\nano\meter} & - & - & - \\ \hline
        $\xi_{F1}^{\text{LR}}$ & \SI{3}{\nano\meter} & - & - & - \\ \hline
        $I_{c,0}^{\text{\text{SR}}}R_N $ & \SI{92.8}{\micro\volt} & - & - & - \\ \hline
        $I_{c,0}^{\text{\text{LR}}}R_N $ & \SI{0.7}{\micro\volt} & - & - & - \\ \hline
\end{tabular}
\caption{Fitting parameters obtained from the fit of Fig. 3 data to Eq. 1, Eq. 2 and Eq. 3.}
\label{tab:pars}
\end{table}

\section{JJ\lowercase{s} with different versions of FM}

\begin{figure}[htbp]
    \includegraphics[width=0.5\textwidth]{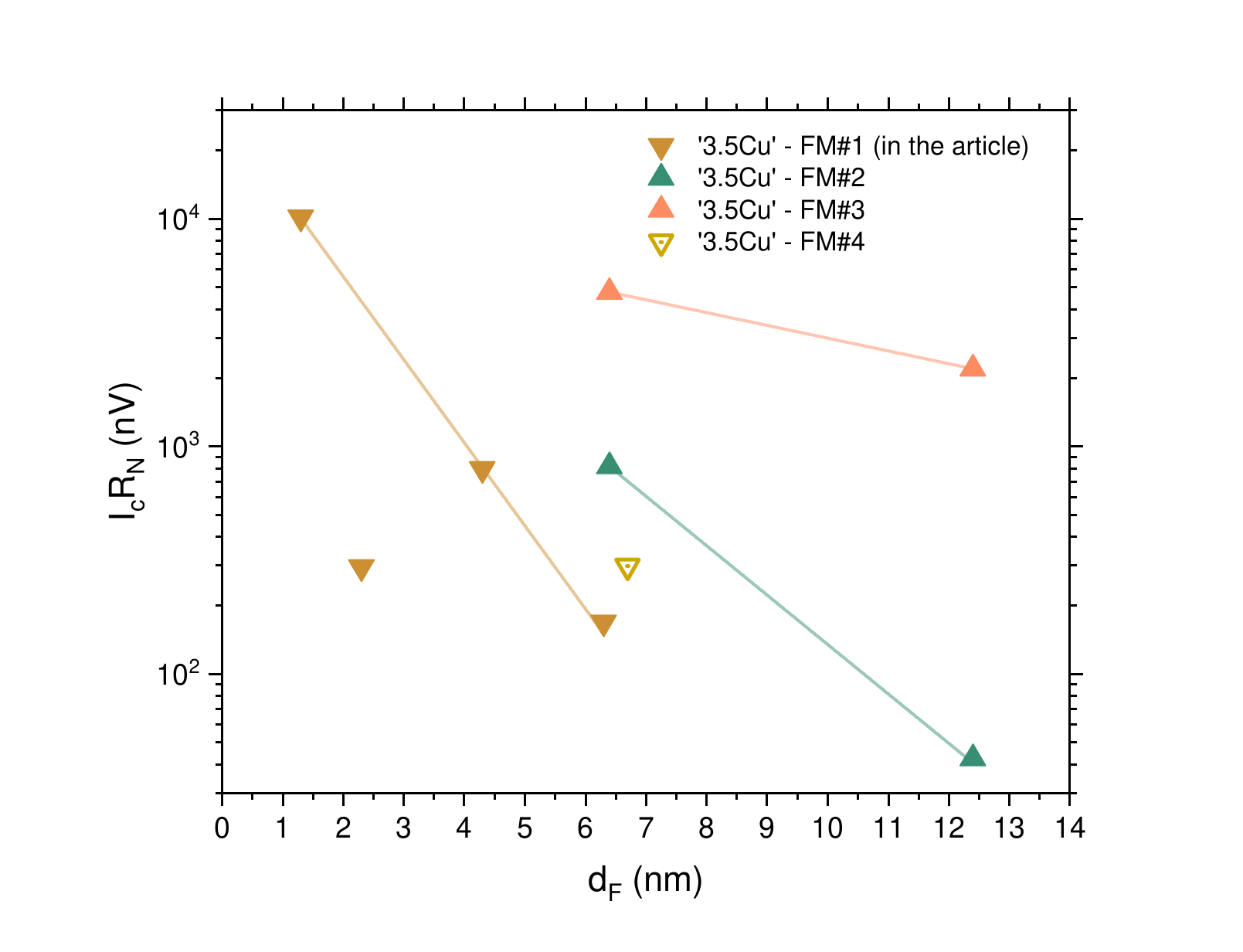}
    \caption{$I_{\text{c}} R_{\text{N}}$ values of JJs with NM = Cu and different FM: FM$\#1$ = same [Co/Ni]$_n$/Co used in the manuscript with Co interfaced to Cu, FM$\#2$ = ``singlet'' weak links from \cite{Gingrich2012Dec}, FM$\#3$ = non-collinear ``triplet'' weak links from \cite{Gingrich2012Dec}, FM$\#4$ = same Co/Ni/Co as FM$\#1$, but Ni is interfaced to Cu instead of Co ([Ni/Co]$_n$/Ni).}
    \label{fig:CuJJ_Extended}
\end{figure}

In Fig. \ref{fig:CuJJ_Extended}, we focus only on NM = Cu JJs. 
FM$\#1$ corresponds to FM weak link used in the main text. $FM\#2$ and $FM\#3$ correspond to [Ni(0.4)/Co(0.2)]$_n$/Ni(0.4) (``singlet'') and Ni(1.2)/Cu(4)/[Ni(0.4)/Co(0.2)]$_n$/Ni(0.4)/Cu(4)/Ni(1.2) (``triplet'') weak links from Ref. \onlinecite{Gingrich2012Dec}, which we reproduce. The estimated decay constants for ``singlet'' and ``triplet'' JJs were found to be $\xi_{F1}^{S}=$\SI{2}{\nano\meter} and $\xi_{F1}^{T}=$\SI{7.8}{\nano\meter}, respectively. We observe that decay in JJs with FM$\#1$ used in the main text is faster than in ``singlet'' JJs with FM$\#2$. We additionally fabricate a ``singlet'' JJs with FM$\#4$, where thicknesses of Co and Ni match the main text, but the Cu is interfaced with Ni instead of Co: [Ni(0.7)/Co(0.3)]$_n$/Ni(0.7). We find supercurrent transmission is lower in JJ with Cu/Co interface than Cu/Ni. We also observe that an increase in the number of Co/Ni interfaces unexpectedly enhanced $I_c R_N$. This result highlights the complexity of identifying LRTC when the LRTC-candidate JJ has interfaces with materials different from the singlet reference. In contrast, the interfaces in singlet JJ and non-collinear triplet JJs are the same but their combination is different.

\section{Verification of the absence of DW nucleation}

\begin{figure}[!ht]
    \includegraphics[width=0.5\textwidth]{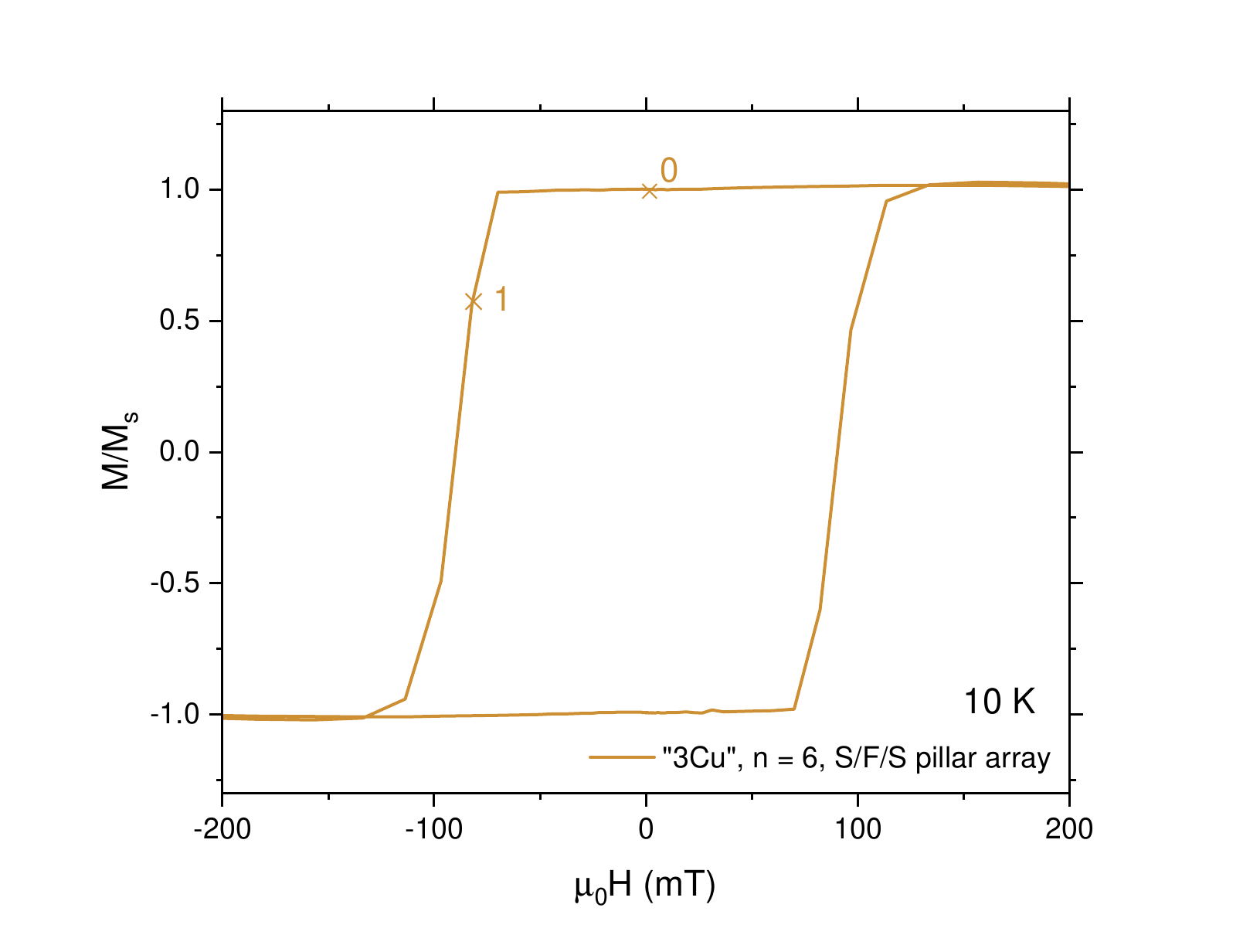}
    \caption{Dependence of magnetic moment on OOP magnetic field measured on the array of S/F/S pillars in `3.5Cu' samples with $n = 6$ ($d_F = \SI{6.3}{\nano\meter}$) JJs}
    \label{fig:Cu6CNC_MH}
\end{figure}

\begin{figure}[!ht]
    \includegraphics[width=0.5\textwidth]{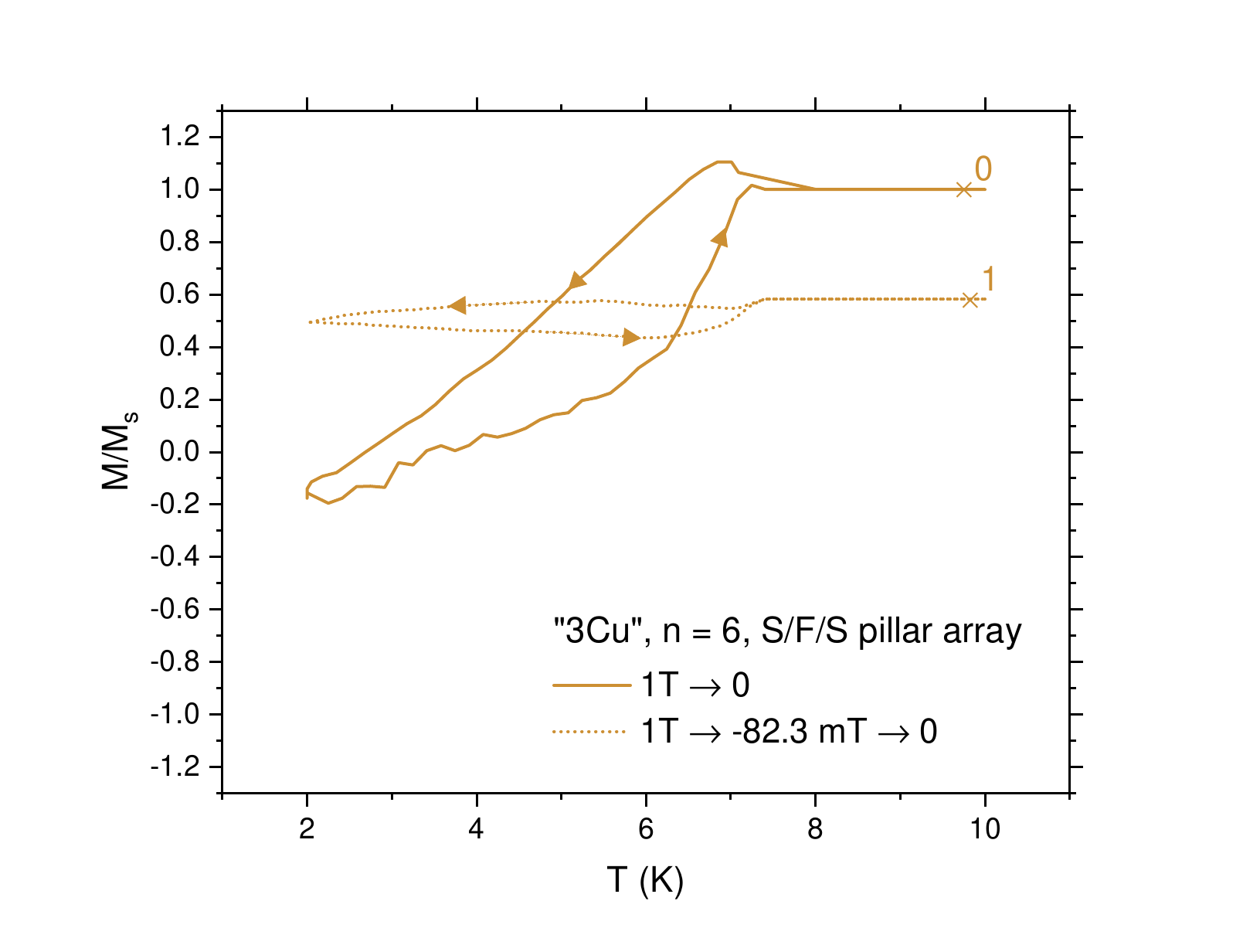}
    \caption{Dependence of magnetic moment on temperature measured on the array of S/F/S pillars corresponding to the composition of  `3.5Cu', $n = 6$ ($d_F = \SI{6.3}{\nano\meter}$) JJs}
    \label{fig:Cu6CNC_MT}
\end{figure}

For this investigation, we deposit extra \SI{60}{\nano\meter} of Nb on top of Si/SiO$_x$/TaN(2)/Nb(50)/Cu(3.5)/ Co(0.7)/[Ni(0.3)/Co(0.7)]$_6$/Cu(3.5)/Nb(10)/Pt(2) film and etch it into an array of \SI{6}{\micro\meter} pillars matching the lateral size of the JJs used in the main text. This fabrication process was done to match magnetic properties of the JJs. First, we carry out $M(H_{\text{OOP}})$ measurement at \SI{10}{\kelvin} using SQUID magnetometry, Fig. \ref{fig:Cu6CNC_MH}. We find values of $M_0$ corresponding to saturated magnetization , and $M_1$ corresponding to intermediate state where domain wall are nucleated. Next, we measure M as a function of T  to examine if magnetic state is affected by superconducting transition of Nb, Fig. \ref{fig:Cu6CNC_MT}. Pillar array is initialized with certain OOP field, cooled down to \SI{2}{\kelvin}, which was the working temperature in the main text, and warmed up to \SI{10}{\kelvin}. In both initial states, $M_0$ and $M_1$, magnetic state was not changed. Therefore, we conclude that spontaneous nucleation of DWs is not the reason for the asymmetric diffraction pattern demonstrated in Fig. 2b in the main manuscript. Alternative cause might be stray field from Abrikosov vortices which may appear in superconducting leads outside of junction area near the JJ rim. Affect of such stray fields has been reported to have \SI{}{\micro\meter} range of influence on Josephson phase. \cite{Golod2019Nov}

\section{Temperature dependence of $I_c$ for `2P\lowercase{t}', $n = 6$ ($d_F = \SI{6.3}{\nano\meter}$) JJ}

\begin{figure}[!ht]
    \includegraphics[width=0.5\textwidth]{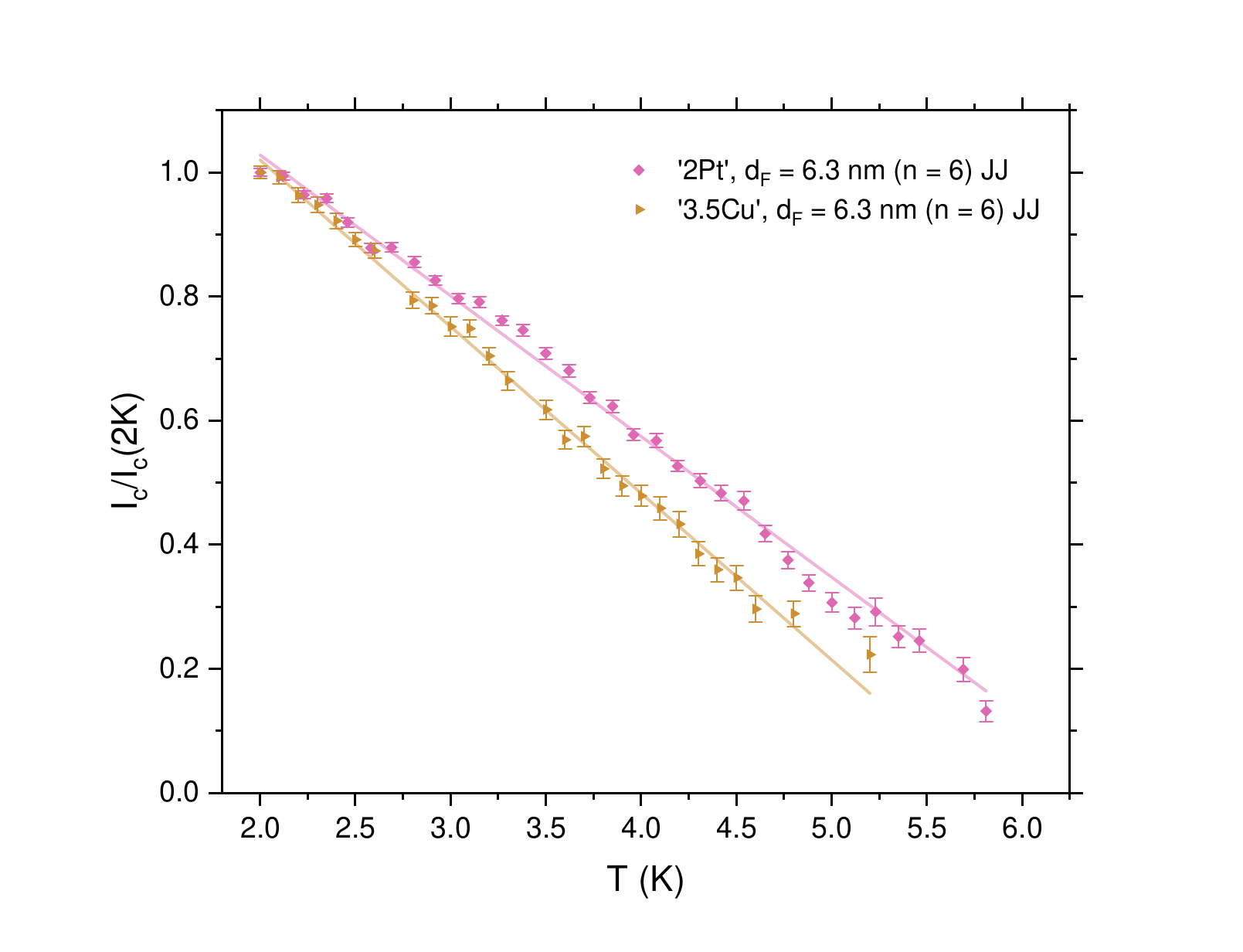}
    \caption{Dependence of normalized critical supercurrent as a function of temperature measured on `2Pt' and `3.5Cu' JJs with $n = 6$ ($d_F = \SI{6.3}{\nano\meter}$). The solid lines are linear fits to the data.}
    \label{fig:IcT}
\end{figure}

Fig. \ref{fig:IcT} shows the dependence of normalized critical supercurrent as a function of temperature measured on `2Pt' and `3.5Cu' with $n = 6$ ($d_F = \SI{6.3}{\nano\meter}$) JJ. The dependencies for both JJs exhibit linear behavior.

\bibliography{references}